\documentclass[]{mn2e}
\usepackage{graphicx}
\usepackage{epsfig}
\usepackage{amsmath}

\newcommand       \K            {\,{\rm K}}

\newcommand       \kabs         {\kappa_{\rm abs}} 

\newcommand	  \kext	        {\kappa_{\rm ext}}
\newcommand	  \ksca	        {\kappa_{\rm sca}}
\newcommand     \gtsim  {\lower.5ex\hbox{$\buildrel > \over \sim$}}
\newcommand     \ltsim  {\lower.5ex\hbox{$\buildrel < \over \sim$}}
\newcommand     \simgt  {\lower.5ex\hbox{$\buildrel > \over \sim$}}
\newcommand     \simlt  {\lower.5ex\hbox{$\buildrel < \over \sim$}}

\newcommand       \mum          {\,{\rm \mu m}}
\newcommand	  \ppm		{\,{\rm ppm}}

\newcommand       \simali       {\sim\,}
\newcommand       \rhom         {\rho_{\rm m}}
\newcommand       \rhoR         {\rho_{\rm R}}

\newcommand	  \md	        {m_{\rm dust}}

\newcommand	  \Cext	        {C_{\rm ext}}
\newcommand	  \Cabs	        {C_{\rm abs}}
\newcommand	  \Csca  	{C_{\rm sca}}

\newcommand	  \aeff	{a_{\rm eff}}

\title{Optical properties of elongated conducting grains}
\author[Huang et al.]
        {X.M.~Huang$^{1,2}$,
        Qi~Li$^{3,2}$\thanks{qili@xtu.edu.cn}, 
        Aigen~Li$^{2}$\thanks{lia@missouri.edu}, 
        J.H.~Chen$^{1,2}$\thanks{jhchen@hunnu.edu.cn},
        F.Z.~Liu$^{4,2}$, and
        C.Y.~Xiao$^{5}$\\
        $^1$College of Physics and Electronics,
               Hunan Normal University, 
               Changsha, Hunan 410081, China\\
        $^2$Department of Physics and Astronomy,
             University of Missouri,
             Columbia, MO 65211, USA\\
        $^3$Hunan Key Laboratory for Stellar
              and Interstellar Physics
              and School of Physics and Optoelectronics,
              Xiangtan University, Hunan 411105, China\\
        $^4$College of Information Science and Engineering,
               Hunan Normal University, 
               Changsha, Hunan 410081, China\\
        $^5$Department of Astronomy,
                Beijing Normal University,
                Beijing 100875, China\\
                }
\begin{document}
\date{}
\pagerange{\pageref{firstpage}--\pageref{lastpage}} \pubyear{2021}

\maketitle

\label{firstpage}
\begin{abstract}
Extremely elongated, conducting dust particles
(also known as metallic ``needles'' or ``whiskers'')
are seen in carbonaceous chondrites and in samples
brought back from the Itokawa asteroid.
Their formation in protostellar nebulae and subsequent
injection into the interstellar medium have been
demonstrated, both experimentally and theoretically.
Metallic needles have been suggested to explain
a wide variety of astrophysical phenomena,
ranging from the mid-infrared interstellar extinction
at $\simali$3--8$\mum$
to the thermalization of starlight to generate 
the cosmic microwave background.
To validate (or invalidate) these suggestions,
an accurate knowledge of the optics
(e.g., the amplitude and the wavelength
dependence of the absorption cross sections)
of metallic needles is crucial.
Here we calculate the absorption cross sections
of iron needles of various aspect ratios 
over a wide wavelength range,
by exploiting the discrete dipole approximation,
the most powerful technique for rigorously calculating
the optics of irregular or nonspherical grains. 
Our calculations support the earlier findings that
the antenna theory and the Rayleigh approximation,
which are often taken to approximate
the optical properties of metallic needles,
are indeed inapplicable.
\end{abstract}

\section{Introduction}\label{sec:intro}
Extremely elongated, needle- or whisker-like metallic grains
are long known to be present in extraterrestrial environments.
Bradley et al.\ (1996) performed an analytical transmission
electron microscopic examination of the Martian meteorite
ALH84001 and reported the detection of nano-sized
magnetite (Fe$_3$O$_4$) whiskers and platelets in association
with carbonates in fracture zones within ALH84001. 
Based on Raman imaging and electron microscopy,
Fries \& Steel (2008) discovered graphite whiskers
in three CV-type carbonaceous chondrites.
They found that the graphite whiskers are associated
with high-temperature calcium-aluminum inclusion
rims and interiors.
Steele et al.\ (2010) conducted two- and three-dimensional
confocal Raman imaging spectroscopy of an Apollo 17
impact melt breccia and found micron-sized graphite,
rolled-graphene sheets and graphite whiskers,
probably resulting from the impact processes responsible
for breccia formation.
More recently, Matsumoto et al.\ (2020) obtained
high-resolution transmission electron microscopic
images of the dust particles returned by
the Japanese Hayabusa mission to the Itokawa asteroid
and reported the discovery of iron whiskers
on asteroidal particles.

Nuth et al.\ (2010) have experimentally shown
that abundant graphite whiskers could be formed
on or from the surfaces of graphite grains 
if they are repeatedly exposed to H$_2$, 
CO, and N$_2$ at 875$\K$, a condition mimicing 
that of protostellar nebulae. 
These newly formed graphite whiskers
could subsequently be expelled
from protostellar systems 
either in polar jets or by radiation pressure 
and populate the interstellar space
(Bland 2008, Nuth et al.\ 2010).
Hoyle \& Wickramasinghe (1988) suggested that
metallic whiskers could form efficiently
in supernova ejecta through screw dislocation.
Piotrowski (1962) argued that electrically charged
interstellar grains, if elongated, tend to grow longer
in the interstellar medium (ISM) through preferential
capture of ions near the ends of the grain. 
In addition, Nuth et al.\ (1994), Nuth \& Wilkinson (1995)
and Marshall et al.\ (2005) have shown both experimentally
and theoretically that the nano-sized iron grains
in the solar nebula could grow to whiskers
with an almost infinite length-to-diameter ratio
at greatly enhanced rates
due to the electrostatic and magnetic dipole interactions.

Metallic whiskers or needles have unique,
elongation-dependent optical properties.
Because of this, they have been invoked to
explain a wide variety of astrophysical phenomena.
Exceedingly elongated metallic needles, 
presumably present in the intergalactic medium (IGM), 
have been suggested as a source of starlight opacity
to thermalize starlight in a steady-state cosmology
to create a non-cosmological microwave background
(Hoyle \& Wickramasinghe 1988), as well as to explain
the microwave background anisotropy (Narlikar et al.\ 2003)
detected by the Wilkinson Microwave Anisotropy Probe
(Bennett et al.\ 2003). Such needles have also been investigated
by Wright (1982) and Aguirre (2000) as a thermalizing agent 
of starlight to generate a cold big bang microwave background.
Aguirre (1999) and Banerjee et al.\ (2000) have also explored
the possibilities of needles as a source of the gray opacity
needed to account for the observed redshift-magnitude relation
of Type Ia supernovae (Riess et al.\ 1998; Perlmutter et al.\ 1999) 
without resorting to a positive cosmological constant. 
More recently, Melia (2020, 2021) has thoroughly
examined the possible rethermalization of
the cosmological microwave background photons
by dust ejected into the intergalactic medium
by the first-generation Population III stars
at redshift $z < 16$. 
In addition, metallic needles have been suggested
by Dwek (2004a) and explored also by Wang et al.\ (2014) 
as an explanation for the anomalously flat mid infrared (IR)
interstellar extinction at $\simali$3--8$\mum$
seen in both diffuse and dense environments
(Lutz 1999, 
Indebetouw et al.\ 2005,
Jiang et al.\ 2006, 
Flaherty et al.\ 2007,
Gao et al.\ 2009,
Nishiyama et al.\ 2009, 
Wang et al.\ 2013,
Xue et al.\ 2016,
Hensley \& Draine 2020),
in stark contrast to the deep minimum expected 
from standard interstellar dust models 
(see Draine 1989, Weingartner \& Draine 2001).
Metallic needles 
have also been suggested by Dwek (2004b)
as the source for the submillimeter excess
observed by Dunne et al.\ (2003)
along the line of sight toward
the Cas~A supernova remnant
(but also see Krause et al.\ 2004,
Wilson \& Batrla 2005,
Gomez et al.\ 2005, 2009).

The ascription of the aforementioned astrophysical
phenomena to metallic needles is largely based on
their conceived unique, anomalously large opacities
over a wide range of wavelengths from the ultraviolet (UV)
all the way to the IR, submillimeter and millimeter,
with the wavelength span depending on their geometrical
(e.g., the length over radius ratio)
and physical (e.g., the resistivity) properties. 
Unfortunately, there lacks exact solution to
the scattering and absorption of light
with metallic needles.
In the literature, the absorption cross sections
of metallic needles are often derived either from
extremely elongated spheroids 
under the Rayleigh approximation
(see Li 2003 and references therein) 
or from the antenna approximation
(see Xiao et al.\ 2020 and references therein).
However, it has been shown that,
for highly conducting needles,
the Rayleigh criterion is not satisfied
and therefore the Rayleigh approximation
is invalid (see Li 2003).
On the other hand, very recently it has also
been shown that the antenna approximation
is not an appropriate representation
for the absorption cross sections of
metallic needles since it violates
the Kramers-Kronig relation
(see Xiao et al.\ 2020).

In this work we aim at calculating
the absorption and scattering cross sections
of metallic needles of various elongations
over a wide wavelength range
from the far-UV to the mid-IR 
by exploying the discrete dipole approximation
(DDA; Purcell \& Pennypacker 1973, Draine 1988),
a powerful technique known to be capable of
accurately calculate the optical properties of
nonspherical or irregular particles. 
In \S\ref{sec:DDA} we briefly present
the computational method and targets.
The results are presented
and discussed in \S\ref{sec:results}.
Our major conclusion is summarized
in \S\ref{sec:summary}.

\begin{figure*}
\centering
\includegraphics[width=.99\textwidth]{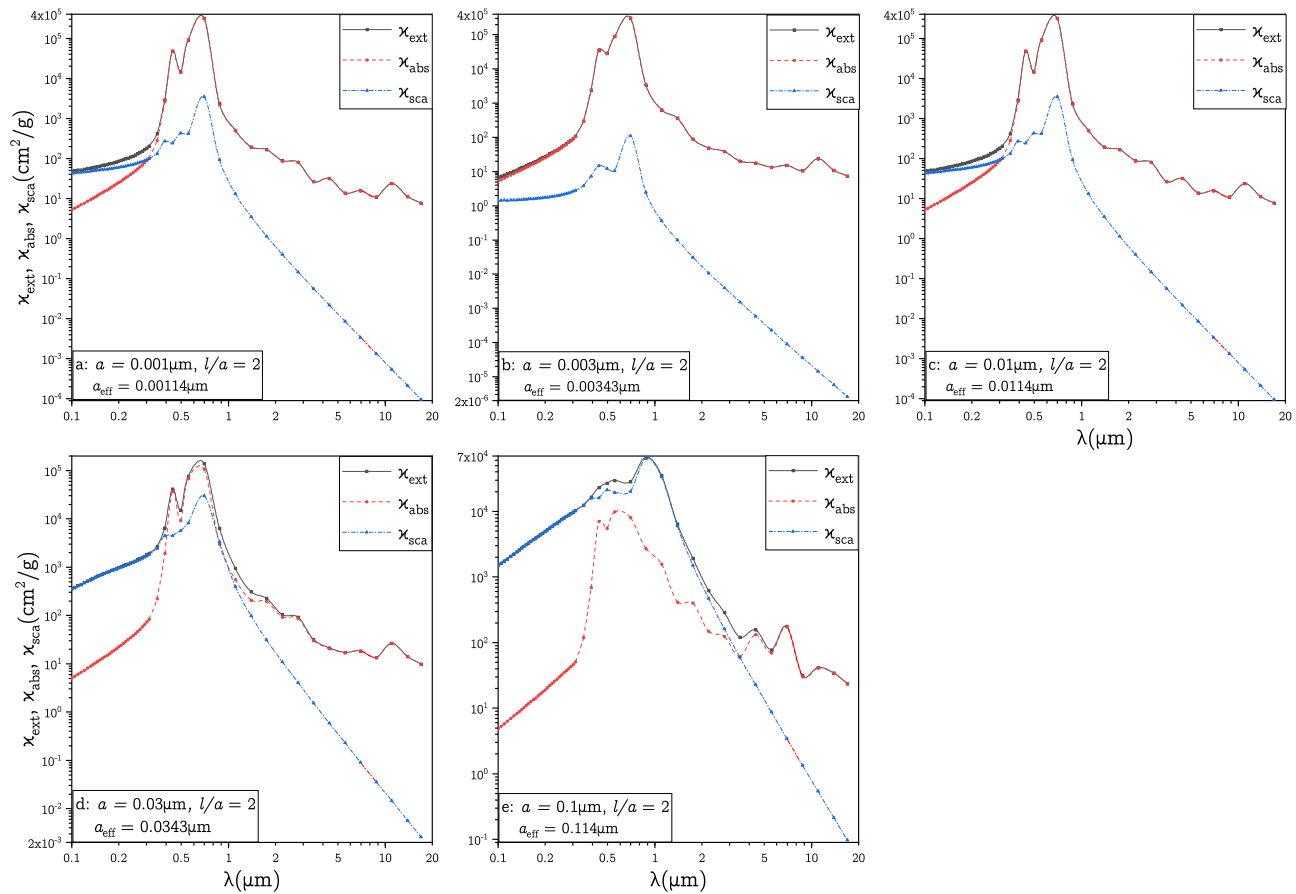}
\caption{
        \label{fig:l2a2}
        Extinction (black solid lines),
        absorption (red dashed lines),
        and scattering (blue dot-dashed lines)
        mass coefficients of circular cylindrical
        iron grains of an elongation $l/a=2$
        and radii $a=0.001\mum$ (a),
        0.003$\mum$ (b), 0.01$\mum$ (c), 
        0.03$\mum$ (d), and 0.1$\mum$ (e). 
        }
\end{figure*}

\begin{figure*}
\centering
\includegraphics[width=.99\textwidth]{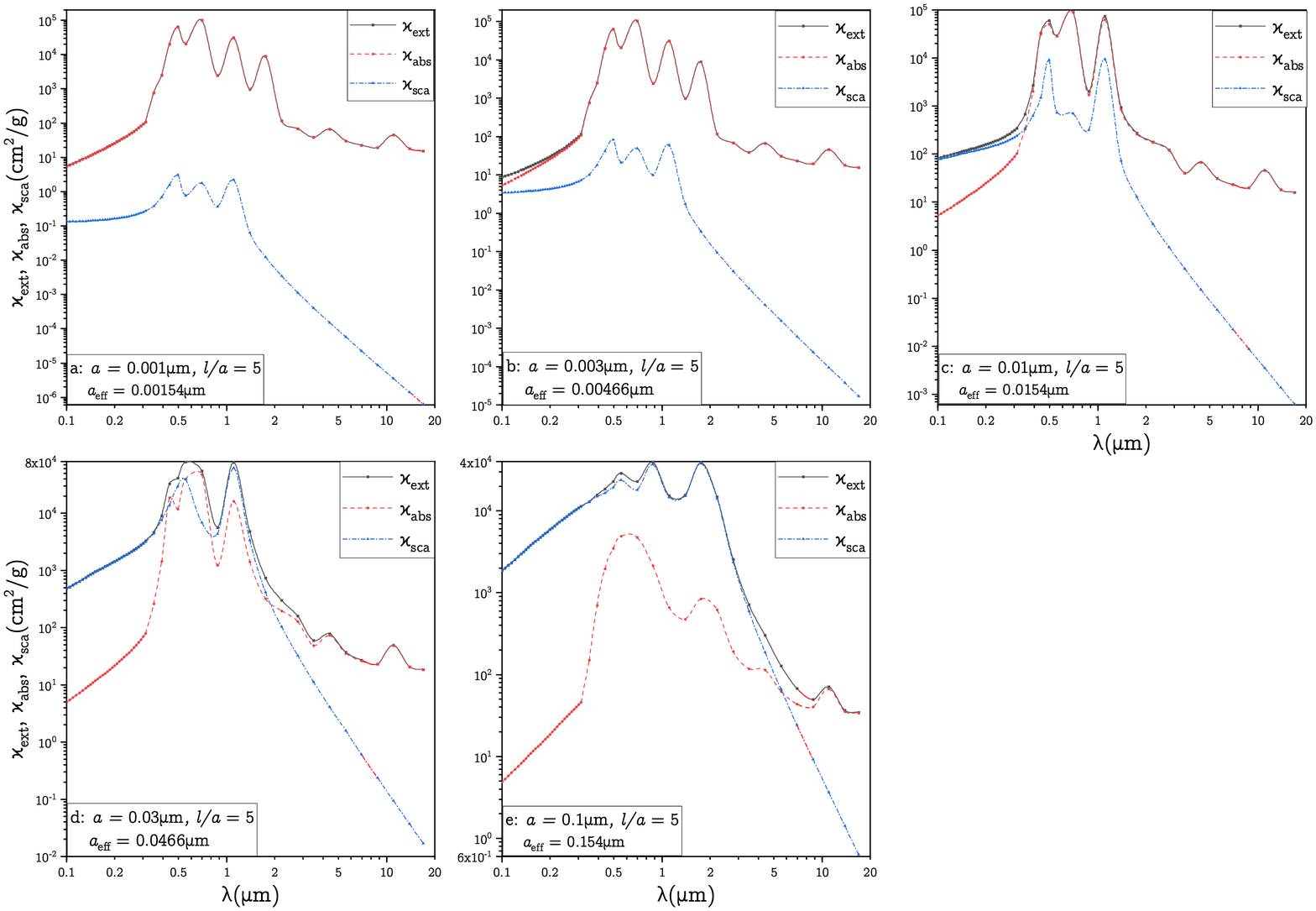}
\caption{
        \label{fig:l2a5}
        Same as Figure~\ref{fig:l2a2} but for
        iron grains of an elongation $l/a=5$.
        }
\end{figure*}

\begin{figure*}
\centering
\includegraphics[width=.99\textwidth]{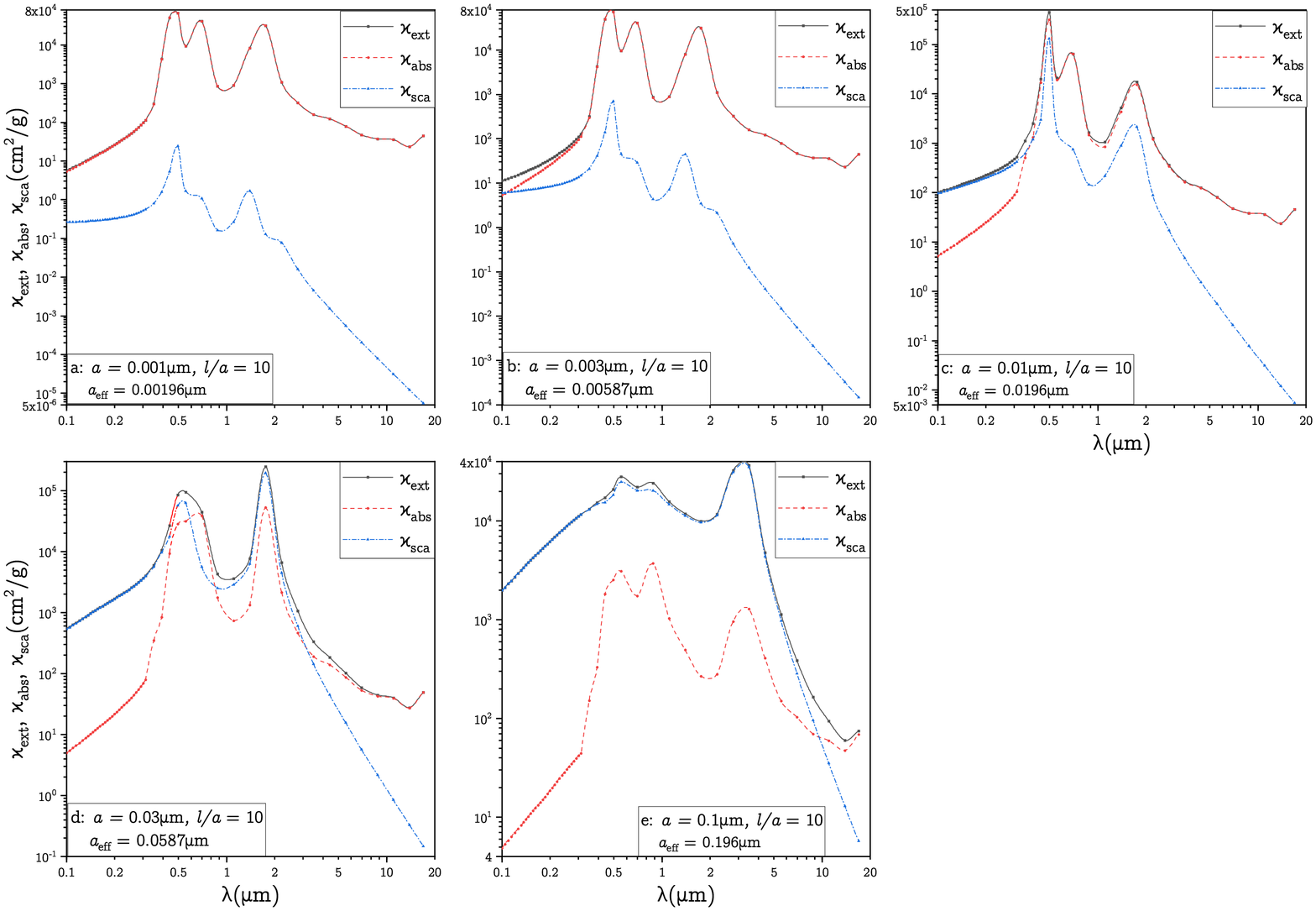}
\caption{
        \label{fig:l2a10}
        Same as Figure~\ref{fig:l2a2} but for
        iron grains of an elongation $l/a=10$.
        }
\end{figure*}

\begin{figure*}
\centering
\includegraphics[width=.99\textwidth]{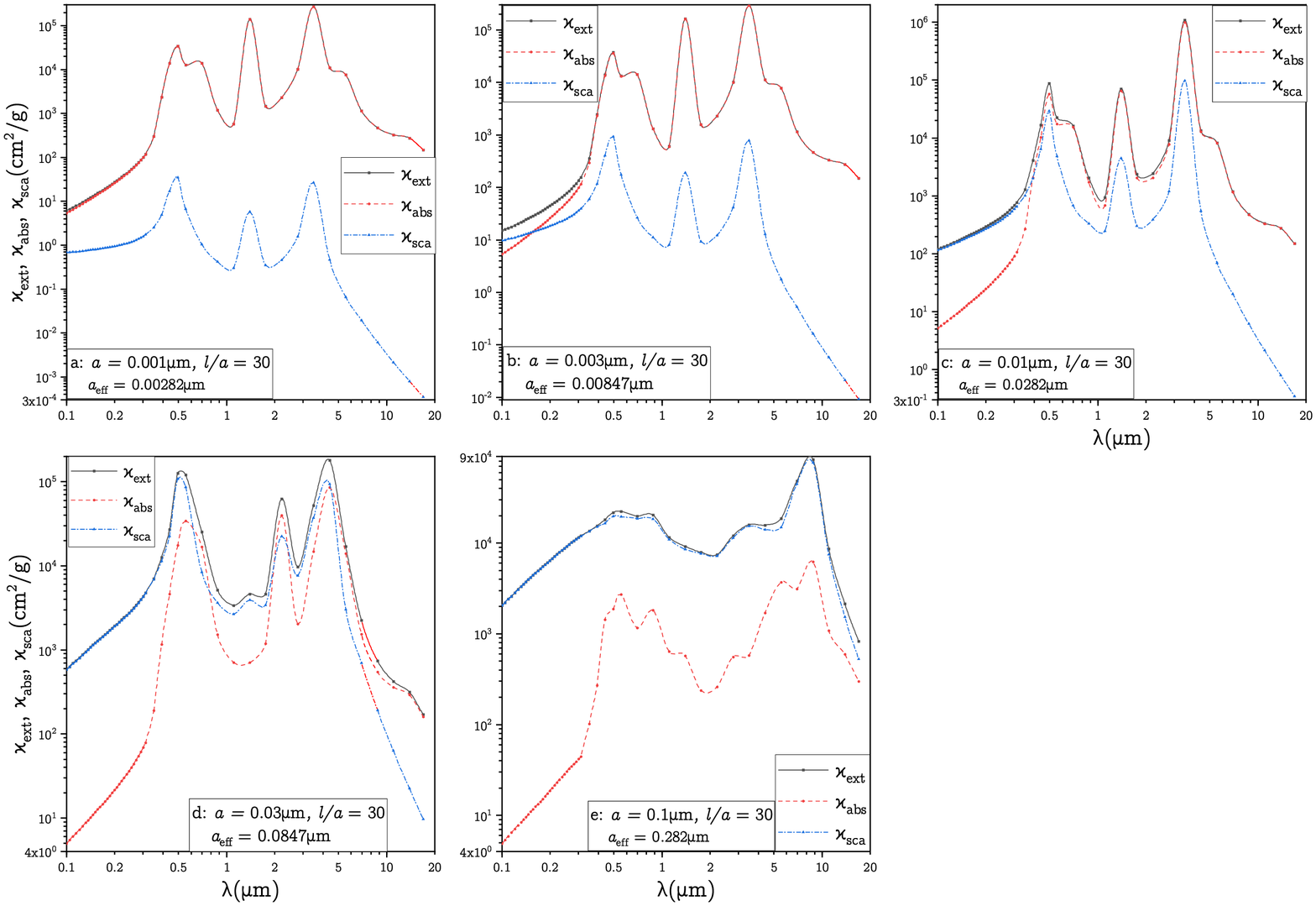}
\caption{
        \label{fig:l2a30}
        Same as Figure~\ref{fig:l2a2} but for
        iron grains of an elongation $l/a=30$.
        }
\end{figure*}

\begin{figure*}
\centering
\includegraphics[width=.99\textwidth]{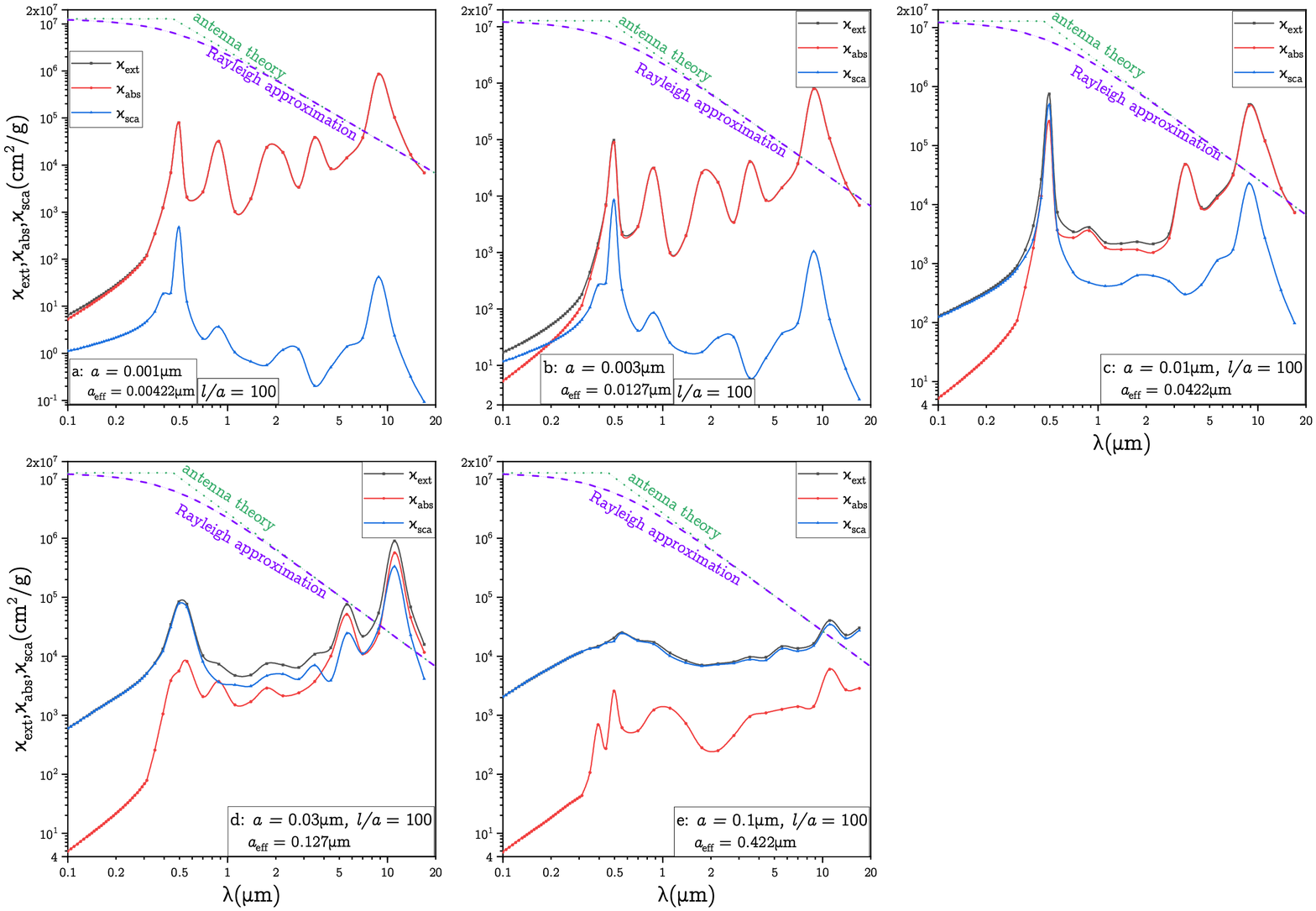}
\caption{
        \label{fig:l2a100}
        Same as Figure~\ref{fig:l2a2} but for
        iron grains of an elongation $l/a=100$.
        Also shown are the extinction mass
        coefficients derived from the Rayleigh
        approximation (purple dot-dot-dashed lines)
        and antenna theory (green dotted lines).
        }
\end{figure*}

\section{The Computational Method}\label{sec:DDA}
The DDA method, also known as
the coupled dipole approximation, 
is a powerful numerical technique
for calculating the scattering and absorption
of electromagnetic radiation by dust particles
of arbitrary geometry.
This method divides a targeted dust particle
into a discrete set of interacting dipoles
on a cubic lattice in which the dipoles are
separated by a distance $d$.
The dipoles, whose properties are defined
by the dielectric properties of the dust material,
can be distributed inside the cubic array
without restrictions and thus,
as long as the inter-dipole lattice spacing $d$
is small enough, the geometry of any
arbitrarily shaped dust particles
can be adequately simulated.
In the cubic array, 
each dipole responds to the incident electric field
as well as the fields contributed by all other dipoles.
%
The collective responses of all these dipoles
allow for the scattering and absorption properties
of dust particles of any geometric shape
and composition to be accurately calculated,
provided that the condition $|m|kd<1$ is satisfied,
where $m$ is the complex refractive index,
$k=2\pi/\lambda$ is the wave number,
and $\lambda$ is the wavelength
(see Draine \&  Flatau 2013).
Due to its versatility and accuracy,
the DDA method has been widely used
to simulate the optical responses of
nonspherical dust particles. 


We perform DDA calculations for elongated conducting grains
by using the DDSCAT7.3 software package
(Draine \& Flatau 2013).
These grains are represented by
circular cylinders of radius $a$, length $l$,
mass density $\rhom$, 
and mass $\md = \pi a^2l\rhom$.
The volume-equivalent sphere radius---the radius
of a sphere having the same volume
as the cylindrical grain 
is $\aeff=\left(3a^2l/4\right)^{1/3}$.
%
We consider iron grains since iron needles
have been invoked to explain a variety of
astrophysical phenomena and there is increasing
evidence for the depletion of Fe in iron grains,
not in silicate dust (see Xiao et al.\ 2020).

\section{Results and Discussion}\label{sec:results}
We consider circular cylindrical iron grains
of five different radii
$a=0.001, 0.003, 0.01, 0.03, 0.1\mum$.
For grains with a given radius, five different
elongations (or aspect ratios) of
$l/a=2, 5, 10, 30, 100$ are considered.
The wavelength-dependent indices of
refraction of iron are taken from Li (2003).
We consider 52 different wavelengths
equally spaced on a logarithmic scale
and ranging from 0.1$\mum$ to 17$\mum$.
These wavelengths cover the far-UV to
the mid-IR range which is relevant to
the optical diagnostic of starlight extinction
and heating by iron grains in the ISM.
Ideally, it would be desirable
if larger grains (up to several submicrons),
larger elongations (up to $\simali$10$^4$),
and longer wavelengths (up to millimeters)
can be considered.
However, in order to meet the criterion
of $|m|kd<1$ required for DDSCAT7.3
to be accurate, the number of dipoles
used to represent such grains would
considerably exceed 10$^{6}$ so that 
the inter-dipole distance $d$ would not
exceed $\lambda/\left(2\pi |m|\right)$.\footnote{%
  Note that for conducting materials like iron,
  the index of refraction $m$ increases with 
  wavelength (see Li 2003) and the required
  number of dipoles would be even larger
  at longer wavelengths.
  }
This poses a severe challenge on
computational demand and makes it impractical.

Let $C_{\rm ext}(a,l,\lambda)$,
$C_{\rm abs}(a,l,\lambda)$,
and $C_{\rm sca}(a,l,\lambda)$
respectively be the extinction,
absorption and scattering cross sections
of a circular iron cylinder of radius $a$
and length $l$ at wavelength $\lambda$.
The extinction, absorption and scattering
mass coefficients of such a grains are
$\kext(a,l,\lambda)\equiv
\Cext(a,l,\lambda)/\left(\pi a^2l\rhom\right)$,
$\kabs(a,l,\lambda)\equiv
\Cabs(a,l,\lambda)/\left(\pi a^2l\rhom\right)$,
and $\ksca(a,l,\lambda)\equiv
\Csca(a,l,\lambda)/\left(\pi a^2l\rhom\right)$,
respectively.

In Figure~\ref{fig:l2a2} we show
the extinction, absorption and scattering
mass coefficients of iron cylinders
of $a=0.001, 0.003, 0.01, 0.03, 0.1\mum$
and $l/a=2$. Most pronounced in the extinction
mass coeffcient profiles is a broad bump
lying in the wavelength range of
$\simali$0.3--1$\mum$.
This broad bump exhibits two resonant peaks
at $\simali$0.44 and $\simali$0.70$\mum$.
These extinction peaks arise from resonances
in the collective motion of free elctrons
constrained to oscillate within small grains
(Bohren \& Huffman 1983).
%

For grains of $a\simlt0.03\mum$, the extinction
is predominantly contributed by absorption,
while scattering is negligible.
As grain size increases, scattering becomes
important and surpasses absorption
for grains with $a=0.1\mum$.
This is probably due to the so-called
``skin depth'' effect, i.e., only a small fraction
of the incident light can ``get into'' the grain
to be absorbed and no appreciable light
penetrates through the grain (Bohren \& Huffman 1983).
Also, the extinction bump is considerably
broadened for grains with $a=0.1\mum$
at the expense of its height.

Similarly, we show in Figures~\ref{fig:l2a5}--\ref{fig:l2a100}
the extinction, absorption and scattering
mass coefficients of iron cylinders
of $l/a=5, 10, 30, 100$.
Again, for each elongation $l/a$,
we consider grains of five different radii
$a=0.001, 0.003, 0.01, 0.03, 0.1\mum$.
Compared with grains of an elongation
$l/a=2$ (see Figure~\ref{fig:l2a2}),
more elongated grains with the same radii
generally tend to exhibit more prominent
resonant structures in their extinction profiles.
These resonant structures are appreciably
broadened and span a wider wavelength range
in grains with larger elongations and larger radii.
Most notably, the sharp extinction peaks
which are prominent in grains of $l/a=100$
and $a\simlt0.03\mum$ fade away in grains
of $l/a=100$ and $a=0.1\mum$.

For highly elongated needle-like conducting grains,
Rayleigh approximation is often employed to estimate
their extinction properties
(see Li 2003 and references therein):
\begin{equation}\label{eq:wright}
  \kext(a,l,\lambda) =
  \left(\frac{4\pi}{3c\rhom\rhoR}\right)
  \left(\frac{1}{1 + \left(\lambda/\lambda_0\right)^2}\right) ~,
\end{equation} 
where $c$ is the speed of light,
$\rhoR$ the dust resistivity, 
and $\lambda_0$, the long-wavelength cutoff,
is estimated from
\begin{equation}\label{eq:cutoff}
\lambda_0 \approx \frac{c\rhoR}{2} 
\frac{\left(l/a\right)^2}{\ln\left(l/a\right)} ~. 
\end{equation}
It is seen that eq.(\ref{eq:cutoff}) establishes
a lower bound on the elongation $l/a$
of the needles which absorb strongly at
wavelengths out to $\lambda_0$.

We have also calculated the extinction mass
coeffcients for iron cylinders of $l/a=100$
and $a=0.001, 0.003, 0.01, 0.03, 0.1\mum$
by making use of the Rayleigh approximation.
As shown in Figure~\ref{fig:l2a100},
it is obvious at a glance that
the extinction mass coeffcients derived from
the Rayleigh approximation are totally different
from that calculated from DDSCAT7.3
which is believed to be accurate.
This demonstrates that the Rayleigh
approximation is not valid for iron needles.
Indeed, Li (2003) has shown that the Rayleigh
approximation is not applicable to conducting needles
since the Rayleigh criterion is not satisfied.

Similarly, conducting needles are often also
approximated as antennae and their extinction
properties are estimated from the antenna theory
(see Xiao et al.\ 2020 and references therein):
\begin{eqnarray}\label{eq:cabs2}
\kext(a,l,\lambda) & = &
\left(\frac{4\pi}{3c\rhom\rhoR}\right)
                         ~~,~~\lambda \le \lambda_0~,\\
&=&
\left(\frac{4\pi}{3c\rhom\rhoR}\right)
\left(\frac{\lambda}{\lambda_0}\right)^{-2}~~,~~\lambda > \lambda_0~,
\end{eqnarray}
where the long-wavelength cutoff
$\lambda_0$ is estimated from eq.\ref{eq:cutoff}.

We have also calculated the extinction mass
coeffcients for iron cylinders of $l/a=100$
and $a=0.001, 0.003, 0.01, 0.03, 0.1\mum$
by making use of the antenna theory.
The results, as shown in Figure~\ref{fig:l2a100},
are close to that of the Rayleigh approximation,
but markedly differ from those computed from
DDSCAT7.3, not only in the extinction magnitudes 
but also in the extinction spectral shapes.
Evidently, neither the Rayleigh approximation
nor the antenna theory can account for
the resonant structures that are conspicuous
in the extinction profiles calculated from DDSCAT7.3.
Although considerably squelched, they are still
perceptible in the extinction profile of grains
of $l/a=100$ and $a=0.1\mum$.
Therefore, like Rayleigh approximation, 
the antenna theory is not an appropriate 
representation for the extinction properties
of metallic needles.
As a matter of fact, Xiao et al.\ (2020) have
shown that the antenna approximation
violates the Kramers-Kronig relation.



It is interesting to note that,
as mentioned earlier,
the resonant structures are notably
damped in the extinction profiles of
grains of larger radii and larger elongations
(e.g., see Figure~\ref{fig:l2a100} for
$l/a=100$ and $a=0.1\mum$),
although far less smooth than
those derived from the antenna theory
and the Rayleigh approximation. 
Such grains could provide the ``gray''
opacity to explain the flat mid-IR
extinction at $\simali$3--8$\mum$
seen in various interstellar regions.
A rough estimate suggests that
iron needles of $l/a=100$ and $a=0.1\mum$,
with Fe/H\,$\simali$20$\ppm$,
are capable of accounting for
the observed mid-IR extinction.

Finally, we also note that the resonant spectral features
seen in the extinction profiles of elongated,
individual grains are unlikely present in
the ISM. First of all, metallic needles,
if indeed present in the ISM, would likely
consist of a distribution of radii and elongations.
It is unlikely that they are just
a string of identically-sized grains
in a perfectly straight line of the same length.
While individual grains exhibit resonant structures
in their extinction profiles which are specific to
their radii and lengths, these fine structures
will be smoothed out if a distribution of
grain sizes are incorporated into
a distribution of chain lengths
(see \S11.3.2 in Bohren \& Huffman 1983)
that would be more likely to represent
the actual grains in the ISM.
Also, actual conducting grains in the ISM
are unlikely as perfectly straight as graphite needles
or rolled graphene sheets
(e.g., see Kimura \& Kaito 2008, Silva et al.\ 2021).
In fact, images of experimentally-grown grain aggregates
(grown either magnetically or electrostatically)
display a wide range in angles between any
three adjacent dust grains
and are ``kinky'' at almost every junction
(see Nuth et al.\ 1994, 2010).
The spectral resonances seen in individual grains
are expected to be cancelled out
in such ``kinky'' aggregates of grains
of a range of sizes, chain lengths
and inter-grain angles
(see Bohren \& Huffman 1983).


\section{Summary}\label{sec:summary}
We have utilized DDA to simulate the interaction
of elongated conducting iron grains of various
sizes and elongations with electromagnetic
waves from the far-UV to the mid-IR.
It is found that their extinction profiles 
exhibt pronounced resonant structures
in the optical and near-IR
which are attributed to resonances
in the collective motion of free elctrons
constrained to oscillate within small grains.
These resonant structures are considerably
broadened and eventually fade away,
if not totally obliterated,
as the grain size and/or elongation increases.
The extinction profiles of needle-like
iron grains of aspect ratios
(i.e., length-to-radius ratios) $\simgt100$
and of radii $a\simgt0.1\mum$
are relatively smooth and flat
in the optical, near- and mid-IR
and could provide the ``gray'' opacity
for accounting for the flat mid-IR
extinction at $\simali$3--8$\mum$
observed in various interstellar regions.
It is also shown that the widely-adopted
Rayleigh approximation and antenna theory
do not provide valid representations
of the optical properties of highly elongated
conducting needle-like grains.
Due to the extremely high computational
demand of DDA, it is presently rather
challenging to calculate the scattering
and absorption properties
of large conducting grains
of high elongations
at long wavelengths.
We call for experimental measurements 
of the optical properties
of conducting needles of various aspect ratios 
over a wide wavelength range
to bound theoretical calculations.

\section*{Acknowledgements}
We thank Drs. B.T.~Draine, E.~Dwek,
J.A.~Nuth III, D.~Pfenniger, and J.L.~Puget 
for stimulating discussions and suggestions.
JHC and XMH are supported in part
by NSFC Grant No.\,U1731107.
AL is supported in part by 
a NSF grant AST-1816411.
CYX is supported in part by 
the Talents Recruiting Program of Beijing Normal University
and the National Natural Science Foundation of China 
(NSFC) Grant No.\,91952111.

\section*{Data Availability}
The data underlying this article will be shared 
on reasonable request to the corresponding authors.

\bsp
\label{lastpage}
\end{document}